\documentclass[11pt]{article} 
\usepackage{hyperref} 
\pdfoutput=1 
 
\begin{document} 
 
\title{Synchronous Droplet Microfluidics:  \\  One ``Clock" to rule them all} 
 
\author{Georgios Katsikis$^1$ and Manu Prakash $^2$\\ 
\\\vspace{6pt} 1. Department of Mechanical Engineering 
\\ 2. Department of Bioengineering
\\ Stanford University, Stanford, CA 94305, USA} 
 
\maketitle 
 
 
\begin{abstract} 
Controlling fluid droplets efficiently in the microscale is of great interest both from a basic science and a technology perspective. We have designed and developed a general-purpose, highly scalable microfluidic control strategy through a single global clock signal that enables synchronous control of arbitrary number of droplets in a planar geometry. A rotating precessive magnetic field provides a global clock signal, enabling simultaneous control of droplet position, velocity and trajectories. Here, in this fluid dynamics video, we explain the main physics of this new microfluidic concept. Video data from droplets moving in sync in different fluidic circuits are included. The experimental setup is described and video data is analyzed to provide a detailed view of the time-dynamics of propagating droplets. Finally, we explore the operational limits of this concept, scaling and phase diagram with physical regime diagram. 

\end{abstract} 
 
 
\section{Introduction} 
 
The {\em hyperref} package is used to make links to the videos. 
 
Two sample videos are 
\href{http://ecommons.library.cornell.edu/bitstream/1813/8237/2/LIFTED_H2_EMS
T_FUEL.mpg}{Video1} and 
\href{http://ecommons.library.cornell.edu/bitstream/1813/8237/4/LIFTED_H2_IEM
_FUEL.mpg}{Video2}. \\
 
The video depicts experimental setup and data for a new method for synchronous microfluidics. The depicted method solves an important problem in droplet microfluidics \cite{prakash}, where a single global clock signal can synchronize movement of an arbitrary number of droplets in planar fluidic circuits. The inspiration for this method comes from an electronic computer memory called Magnetic Bubble Memory \cite{romankiw}. We depict some simple test circuits (droplet loops, interconnected loops) and show data \cite{katsikis} for trajectory of droplets via computational tracking. The video explores limits of synchronous propagation and operational limits. Range of bias magnetic field and rotating fields is displayed during the videos. Driving rotating field characterized by operational frequency determines the effective average propagation velocity of a droplet and hence the Reynolds number of the system. Videos shown here were collected using a Phantom V5 high speed camera at various frame rates.\\ 

We acknowledge valuable advice from all members of Prakash Lab at Stanford University. A patent has been filed for methods described in these videos by Stanford OTL. The authors declare no competing interests.

\end{document}